\documentclass[12pt]{article}
\usepackage{epsfig}
\newcommand{\PL}[3]{Phys. Lett. {\bf {#1}} ({#2}) {#3}}

\newcommand{\PR}[3]{Phys. Rep. {\bf {#1}} ({#2}) {#3}}
\newcommand{\NP}[3]{Nucl. Phys. {\bf {#1}} ({#2}) {#3}}
\newcommand{\PRL}[3]{Phys. Rev. Lett. {\bf {#1}} ({#2}) {#3}}

\newcommand{\ga}{\gamma_0}
\newcommand{\am}{\frac{\gamma_0}{q_1}}
\newcommand{\an}{\frac{\gamma_0}{q_2}}
\newcommand{\amn}{\frac{\gamma_0}{q_1+q_2}}
\newcommand{\bm}{\gamma_0  \,q_1}
\newcommand{\bn}{\gamma_0  \,q_2}
\newcommand{\bmn}{\gamma_0 \,(q_1+q_2)}
\newcommand{\pa}{\partial}

\newcommand{\eq}{\begin{equation}}
\newcommand{\eqx}{\end{equation}}
\newcommand{\eqn}{\begin{eqnarray}}
\newcommand{\eqnx}{\end{eqnarray}}
\newcommand{\Dt}{\Delta}
\newcommand{\dt}{\delta}
\newcommand{\Th}{\theta}
\newcommand{\om}{\omega}
\newcommand{\cor}[1]{\left\langle{#1}\right\rangle}

\newcommand{\paus}{\,\,\,\,\,\,\,\,\,\,\,\,\,\,\,\,\,\,\,\,\,}
\begin{document}
\mbox{ }
\vspace{20mm}

\begin{flushright}
\begin{minipage}[t]{60mm}
{\bf SACLAY-T00-182,\\
    TSL/ISV-00-0239 \\
    December 2000} 
\end{minipage}
\end{flushright}

\begin{center}
\vspace{5mm}

{\bf \Large {\bf Factorial correlators: angular scaling within\\ QCD jets }}

\vspace{20mm}

{\Large R.~Peschanski $^{*}$} \footnote{e-mail:pesch@spht.saclay.cea.fr},  
{\Large B.~Ziaja} $^{**\,\dag}$ \footnote{e-mail:beataz@solaris.ifj.edu.pl}\\  
\vspace{3mm}
         $^{*}$  \it CEA, Service de Physique Th\'eorique, CE-Saclay,\\
                 \it F-91191 Gif-sur-Yvette Cedex, France\\
\vspace{3mm}
		 
         $^{**}$ \it Department of Theoretical Physics,\\
                 \it H.~Niewodnicza\'nski Institute of Nuclear Physics,\\
                 \it 31-342 Cracow, Poland\\ 
\vspace{3mm}
         $^\dag$ \it High Energy Physics,\\
                 \it Uppsala University,\\
                 \it P.O. Box 535, S-75121 Uppsala, Sweden
\end{center}

\vspace{5mm}

{\bf Abstract:} 
Factorial correlators measure the amount of dynamical correlation in  
multiplicity between two separated phase-space windows. We  present the 
analytical derivation of factorial correlators  for a QCD jet described  at the 
double logarithmic  (DL) accuracy. We obtain a new angular scaling property for 
properly  normalized correlators between  two solid-angle cells or two rings 
around the jet axis. Normalized QCD factorial correlators scale with the 
angular distance and are independent of the window size. Scaling violations 
are expected beyond DL approximation, in particular from the subjet structure. 
Experimental tests are feasible, and thus  welcome.
\vfill
\clearpage
\setcounter{page}{1}
%%%%%%%%%%%%%%%%%%%%%%%%%%%%%%%%%%%%%%%%%%%%%%%%%%%%%%%%%%%%%%%%%%%%%%%%%%%%%
%%%%%%%%%%%%%%%%%%%%%%%%%%%%%%%%%%%%%%%%%%%%%%%%%%%%%%%%%%%%%%%%%%
\section{Introduction}

Large multiplicity fluctuations observed in high energy  collisions have 
been already studied for many years \cite{l1}. Advanced methods of data
analysis like e.\ g.\ factorial moment approach \cite{l2,l22,l3} have been 
introduced and implemented for the analysis of multiplicity patterns. 
Finally, they led to discovery of {\it intermittency} in multiparticle  
production which refers to the scaling of factorial moments with the size of a 
single bin within the analized pattern \cite{l22,l3}.

Many different models have been proposed for the explanation of 
the effect \cite{l4}. Some of them suggested that an underlying final state 
multiparticle cascade may be responsible for the scaling of particle moments
\cite{l22,l5}. Straightforward calculations performed for 
multiplicative random cascading models \cite{l2,l22,l6} led to the qualitative 
predictions for the scaling behaviour of factorial moments which were 
backed afterwards by analyses proposed in the framework of the standard 
theory of strong interactions (QCD) \cite{l7}. Monte-Carlo simulations 
based on conventional QCD parton cascading tend to describe quite well 
the effect \cite{l8}, confirming the relevance of  scaling properties in QCD 
parton cascading. However, discrepancies in the precise comparison with 
{\it analytical} predictions remain \cite{l8}. 

So far, both the phenomenological and theoretical investigations 
of multiplicity patterns have concentrated mostly on different kinds 
of particle moments estimated for a single bin \cite{l4}. There are, however,
still intriguing questions remaining about the properties of  correlations
between different bins. The observables related to these correlations are
expected to reflect the presence of large dynamical fluctuations underlying
the pattern stronger than the averaged observables estimated for a single bin.
To investigate these bin-bin correlations the factorial correlators 
\cite{l22,l3}
have been introduced.

Factorial correlators seem to contain some extra information on 
multiplicity fluctuations which may be used to complete and adjust the 
information obtained on that from the standard factorial moment analysis 
\cite{l11}. Moreover, the present status of experimental investigations 
\cite{l8}, \cite{l12} allows one to expect that the comparison of model 
predictions
with the real data will be possible soon. It could help to investigate more 
systematically the validity of QCD Monte-Carlo for the  description of 
fluctuations in jets and discuss, using a wider set of data, the relevance and 
problems of analytical QCD calculations.

However, factorial correlators have been studied only in the framework of
phenomenological models \cite{l4}. The rigorous analysis in the framework of
QCD (even at leading log orders) has not been performed so far.

This paper aims to fill the gap by presenting the analytical derivation
of factorial correlators performed for the QCD parton cascade \cite{l7}
at the double logarithmic (DL) accuracy \cite{l14}. For simplicity we consider
only the fixed $\alpha_S$ case, expecting that it gives good qualitative
estimation of scaling exponents as it was realized previously for factorial 
moments case \cite{l7}.
The obtained scaling dependence of the correlators on the relative distance
between the two solid-angle cells recovers the similar result obtained
in the framework of random cascading $\alpha-$model \cite{l2}. This seems
to be a kind of universal relation.

The paper is organized as follows. In the next section we introduce factorial
correlators defined for small solid angle cells around the subjet direction.
In section 3, using the DL generating functional \cite{l14}, we derive the 
inclusive two gluon
distribution which is neccessary to evaluate the correlators. In section 4
the leading contribution to factorial correlator is estimated, and found
to obey a scaling law similar as in ref.\ \cite{l2} for random multiplicative 
cascading models. In section 5 we discuss briefly the modifications which 
could come from relaxing some of our approximations: including
running $\alpha_S$ or energy momentum conservation. Finally, in section 6 
we sum up our results and present our conclusions, including suggestions 
for experimental evaluation of the {\it normalized factorial correlators} 
and a discussion of the new QCD scaling law found in the DL approximation 
of QCD.
%%%%%%%%%%%%%%%%%%%%%%%%%%%%%%%%%%%%%%%%%%%%%%%%%%%%%%%%%%%%%%%%%%%%%%%%%%%%%
%%%%%%%%%%%%%%%%%%%%%%%%%%%%%%%%%%%%%%%%%%%%%%%%%%%%%%%%%%%%%%%%%%%%%%%%%%%%%

\section{Factorial correlators in QCD jets}

%FIGURE 1
\begin{figure}[t]
    \centerline{\epsfig{figure=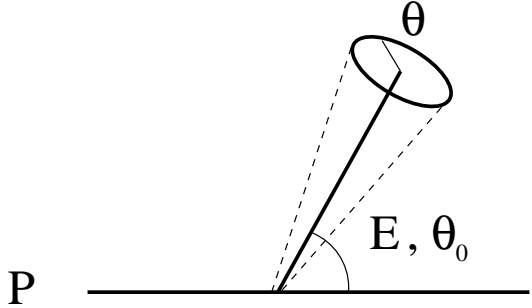,width=7cm}}
    \caption{{\footnotesize 
Example of phase-space cell for the QCD parton cascade. The two-dimensional
cone of half-opening angle $\Th$ is placed at both  solid angle $\Th_0$ 
and azimuthal angle $\phi$ taken with respect to the main jet axis. 
}}
\label{fig.1}
\end{figure}

Normalized factorial moments $F_q$ \cite{l2,l22,l3,l4} designed to study 
multiplicity 
fluctuations in a given phase-space cell of size $\dt$ are defined as~:
\eq
F_q(\dt)=\frac{{\cor {n(n-1)\ldots (n-q+1)}}_{\dt}}{{\cor n}^q_{\dt}},
\label{fq}
\eqx
where $n$ is particle multiplicity in the phase-space cell, and 
the average $\cor{}$ is made over events. Among other types of fluctuations 
studied using factorial moments, the intermittency regime corresponds 
to  moments which  scale with the size of phase-space cell as~:
\eq
F_q(\dt) \sim \left(\frac{\Dt}{\dt}\right)^{\phi_q},
\label{fqscal}
\eqx
where $\Dt$ denotes the size of whole available phase-space, and 
$\phi_q$ is a positive scaling exponent known also as intermittency 
exponent.

In order to study correlations between different phase-space cells one
introduces factorial correlators $F_{q_1,\,q_2}$ (known also as multivariate
factorial moments) \cite{l2,l4} 
which may be regarded as multidimensional extension of 
moments (\ref{fq}). They take the form~:
\eq
F_{q_1,\,q_2}(\dt_1,\dt_2)=
\frac{{\cor {n(n-1)\ldots (n-q_1+1)\!\mid_{\dt_1}\,\,n(n-1)\ldots 
(n-q_2+1)
\!\mid_{\dt_2}}}}{{\cor n}^{q_1}_{\dt_1}\ {\cor 
n}^{q_2}_{\dt_2}},
\label{fq12}
\eqx
where $\dt_1$ and $\dt_2$ denote the sizes of two separate phase-space cells.
Assuming a multiplicative random cascade underlying the particle production,
one predicts a recursive relation between the scaling exponents for
factorial moments and factorial correlators~:
\eq
\phi_{q_1,\,q_2}=\phi_{q_1+q_2}-\phi_{q_1}-\phi_{q_2},
\label{expo}
\eqx
where $\phi_{q_1,\,q_2}$ is the intermittency exponent defined for doubly 
normalized factorial correlators which correspond to factorial correlator 
(\ref{fq12}) divided by factorial moments derived for $\dt_1$ and $\dt_2$ 
cells respectively~:
\eq
\frac{F_{q_1,\,q_2}(\dt_1,\dt_2)}{F_{q_1}(\dt_1)\, 
F_{q_2}(\dt_2)} \sim
\left(\frac{\Dt}{\dt_{12}}\right)^{\phi_{q_1,\,q_2}},
\label{fq12scal}
\eqx
where $\dt_{12}$ is the relative distance between the two phase-space cells
$\dt_1$ and $\dt_2$. Note the interesting feature that the dependence on the 
individual  phase-space cells $\dt_1$ and $\dt_2$ disappears from the 
factorial correlator when they are normalized as in (\ref{fq12scal}).

For the QCD parton cascade \cite{l7} the phase-space cell is more conveniently 
choosen to correspond to a window in emission solid angle $\Th$ (see e.g.
Fig.\ 1). 
The window size $\Th$ is then compared to large scale $\Th_0$ which denotes 
the jet emission angle.
The window may be either one-dimensional ring of aperture $\Th$ placed at the
angle $\Th_0$ with respect to the sphericity axis, or it may be a
two-dimensional cone of half-opening angle $\Th$ placed at both the solid
angle $\Th_0$ and azimuthal angle $\phi$ taken with respect to the main jet
axis.

It was found out that for QCD angular factorial moments \cite{l7} 
there is a scaling relation~:
\eq
F_q(\Th)\sim \left(\frac{\Th_0}{\Th}\right)^{\phi^{QCD}_q},
\label{scalfac}
\eqx
where the intermittency exponent $\phi^{QCD}_q$ calculated for simplicity
in double logarithmic approximation with fixed coupling constant 
$\alpha_S$ reads~:
\eq
\phi^{QCD}_q=\frac{\gamma_0}{q}- \gamma_0\,\,q\,\,+D(q-1).
\label{phiqcd}
\eqx
Number $D$ denotes the window dimension, and it equals $D=1$ for the ring, 
and $D=2$ for the cone. The coefficient $\gamma_0$ is the QCD anomalous
dimension for gluon cascade which for fixed $\alpha_S$~ equals 
$\gamma_0^2=4N_C\frac{\alpha_S}{2\pi}$ \cite{l14}, where $N_C$ denotes 
the number of  colours.

In order to extend the intermittent analysis by investigating also the
possible correlations between particle flows measured at two different
rings around the subjet axis $(E,\,\Th_0)$ let us  introduce the {\it angular 
factorial correlators} for the QCD parton cascade, defined as follows. 
Similarly as for the angular factorial moments one would identify 
the large scale (size of whole available phase-space $\Dt$) with the respective 
subjet emission angle $\Th_{0}$, and the small scales 
$\dt_1$, $\dt_2$ with the window apertures $\Th_1$, $\Th_2$ (cf.\ Fig.\ 2).
We will consider parton flows emitted into two rings
placed at the separation angle $\Th_{12}$ with respect to the subjet axis.
The ring openings are $\Th_1$ and $\Th_2$ respectively (see Fig.\ 2). 
 In using DL approximation framework, we have to assume that the angles are 
small with respect to the subjet direction. More precisely,  
we will assume that they obey the inequalities~:
\eq
\Th_1, \Th_2 \ll \Th_{12} \ll \Th_0.
\label{ineq}
\eqx
Henceforth  $\Th_{01} \sim \Th_{02},$ and the relative bin distance
$\dt_{12}$ in the one-dimensional approximation then corresponds to the 
angular distance $\Th_{12}=\Th_{02}-\Th_{01}$ 
between the two rings. We will discuss the relevance of this DL approximation 
in our discussion in section 5.

%FIGURE 2
\begin{figure}[t]
    \centerline{\epsfig{figure=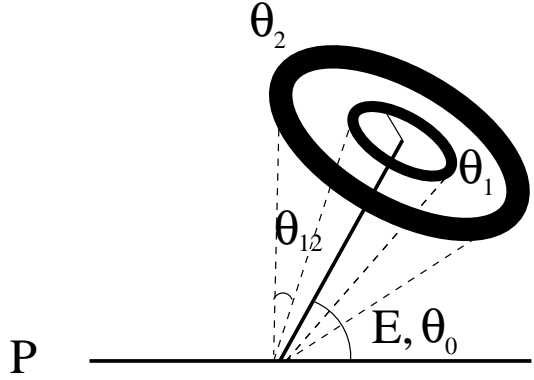,width=7cm}}
    \caption{{\footnotesize 
Phase-space cells for angular factorial correlators. Parton flows are emitted 
into two rings placed at the separation angle $\Th_{12}$ with respect to the 
subjet axis. The ring openings are $\Th_1$ and $\Th_2$ respectively. 
}}
\label{fig.2}
\end{figure}

Having defined angular factorial correlators, we may now estimate them
with a good accuracy from the convolution of the inclusive two-particle 
density \\ 
$D^{(2)}(P;\,E,\Th_0; \,k_1,k_2,\Th_{12},\Th_1,\Th_{2})$ 
with the respective multiplicity moments in the phase-space cells
$\theta_1$ and $\theta_2$. Using (as for the case of factorial moments 
\cite{l7}) 
their expression in the so-called KNO limit proportional up to constants to  
$q_1$th and $q_2$th power of mean multiplicities $N(k_1\,\Th_1)$, 
$N(k_2\,\Th_2)$, we obtain~:
{\footnotesize
\eqn
F_{q_1\,q_2}(\Th_{0};\,\Th_{12},\Th_1,\Th_2)\propto 
\int^{E} \frac{dk_1}{k_1}\,\int^{E} \frac{dk_2}{k_2}\,
D^{(2)}(P;\,E,\Th_0;\,k_1,k_2,\Th_{12},\Th_1,\Th_{2})\, 
N^{q_1}(k_1\Th_1)\,N^{q_2}(k_2\Th_2) \ ,
\label{fcqcd}
\eqnx
}
where $E$ denotes the energy of the subjet.

The mean multiplicity for QCD parton cascade is  dependent on an infrared 
cut-off $\mu$, and it reads~:
\eq
N(k\,\Th)\,\sim\,e^{\gamma_0\ln\frac{k\,\Th}{\mu}}.
\label{meanm}
\eqx
However, similarly as for the factorial moment case we expect that the
cut-off dependence will disappear after normalization, i.\ e.\ when coming to 
normalized factorial  correlators (\ref{fq12scal}). The inclusive 
two-particle density 
$D^{(2)}(P;\,E,\Th_0;\,k_1,k_2,\Th_{12},\Th_1,\Th_{2})$ remains thus
the only unknown quantity, necessary to evaluate (\ref{fcqcd}). 
Its explicit form will be derived in the next section.
%%%%%%%%%%%%%%%%%%%%%%%%%%%%%%%%%%%%%%%%%%%%%%%%%%%%%%%%%%%%%%%%%%%%%%%%%%%%%%
%%
%%%%%%%%%%%%%%%%%%%%%%%%%%%%%%%%%%%%%%%%%%%%%%%%%%%%%%%%%%%%%%%%%%%%%%%%%%%%%%
%%

\section{Inclusive two-particle density}

%
%FIGURE 3
\begin{figure}[t]
    \centerline{\epsfig{figure=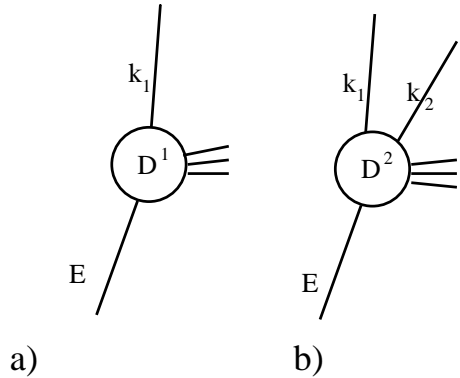,width=6cm}}
    \caption{{\footnotesize 
Diagrammatic representation of the single particle inclusive density 
$D^{(1)}(E/k_1,\Th_{0}/\Th_1)$ (Fig.\ 3a) and the modified inclusive 
two particle density
$D^{(2)}(E/k_1,E/k_2,\Th_{0}/\Th_{1}, \Th_{0}/\Th_{2};\Th_{12})$ (Fig.\ 3b).
}}
\label{fig.4}
\end{figure}
%
%FIGURE 4
\begin{figure}[t]
\centerline{\epsfig{figure=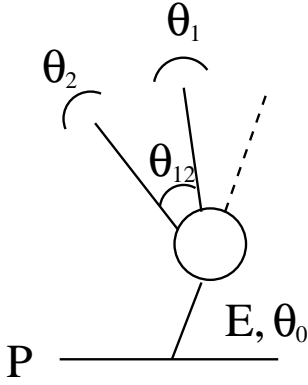,width=4cm}}
\caption{{\footnotesize 
Diagrammatic representation of the convolution 
$D^{(1,ex)}_P(E,\Th_{0})\times D^{(2)}
(E/k_1,E/k_2,\Th_{0}/\Th_{1},\Th_{0}/\Th_{2};\Th_{12})$.
It corresponds to the emission of particles from one subjet $(E,\Th_{0})$ 
originating from the main jet $(P,\Th_P)$ into two rings of apertures  
$\Th_{1}, \Th_{2}$ with separation angle $\Th_{12}$, placed around the 
subjet axis.
}}
\label{fig.3}
\end{figure}

In order to obtain the explicit form of two particle density 
{\footnotesize $D^{(2)}(P;\,E,\Th_0;\,k_1,k_2,\Th_{12},\Th_1,\Th_{2})$}
to insert it into 
(\ref{fcqcd}) we start with a derivation of a related quantity 
$D^{(2)}_p(k_1,k_2)$ from the QCD parton cascade formalism.

The inclusive two-particle density $D^{(2)}_p(k_1,k_2)$ i.e. the inclusive 
density to produce two particles of momenta $k_1$, $k_2$ from a parent 
particle of momentum $p$ may be derived in a convenient way from the 
generating functional for the  QCD parton cascade \cite{l14}.
This functional in DL approximation with fixed $\alpha_S$ takes the form~:
\eq
Z_p[u]\,=\,u(p)\,
e^{\int d^3k\,M_p(k)\,(Z_k[u]-1)}
\label{genfun}
\eqx
with the initial condition $Z_p[u]\mid_{\{u=1\}}\,=\,1$. The function $u(p)$
is a probing function while the  factor $M_p(k)$ describes the DL probability of
emitting a particle of momentum $k$ from a primary particle of momentum $p$.
It reads~:
\eq
d^3k\,M_p(k)\,=\,\gamma_0^2\,\frac{d\Th}{\Th}\,\frac{dk}{k}\frac{
\phi}{2\pi}
{\bar \Th}_p(k),
\label{mpk}
\eqx
where cut-off $\Th$-function ${\bar \Th}_p(k)$  contains phase-space limitations
resulting from possible angular and energy ordering between parent
particle $p$ and child particle $k$~:
\eq
{\bar \Th}_p(k)=\left\{ p>k, \,\,\Th_{pk}<\Th_p, 
\,\,k\Th_{pk}>\mu \right\},
\label{teta}
\eqx
where $\mu$ is the infrared cut-off.

The inclusive two-particle density is then defined as a 
functional derivative of $Z_p[u]$~:
\eq
D^{(2)}_p(k_1,k_2)\,=\,k_1\,k_2\,\frac{\dt^2\,Z_p}{\dt u(k_1)\, 
\dt u(k_2)}
\vert_{\{u=1\}},
\label{d2}
\eqx
which results in a recursive equation for $D^{(2)}_p(k_1,k_2)$~:
\eq
D^{(2)}_p(k_1,k_2)=D^{(1)}_p(k_1)\,D^{(2)}_p(k_2)\,
-\,\dt(\ln(p/k_1))\,\dt(\ln(p/k_2))\,
+\,\int\,d^3k\,M_p(k)\,D^{(2)}_k(k_1,k_2),
\label{d2e}
\eqx
where $D^{(1)}_p(k_1)\,D^{(2)}_p(k_2)\,$  denote the single particle inclusive 
densities for particles $k_1$ and $k_2$ respectively, and the integration
limits are defined by (\ref{teta}) ($max(\Th_1,\Th_2)<\Th<\Th_p$,
$max(k_1,k_2)<k<p$).

Taking into account the scaling properties of the DL phase-space measure 
(\ref{mpk}), one obtains an equation for the modified inclusive 
two-particle density\\
$D^{(2)}(E/k_1,E/k_2,\Th_{0}/\Th_1,\Th_{0}/\Th_2;\Th_{12})$
with  a scaling dependence on the relevant variables 
(choosen here to be $p\equiv E,$ the energy of the subjet, and $k_1>k_2$, 
$\Th_1<\Th_2$)~:
{\footnotesize
\eqn
D^{(2)}_p(k_1,k_2)\Rightarrow 
D^{(2)}(E/k_1,E/k_2,\Th_{0}/\Th_1,\Th_{0}/\Th_2;\Th_{12})=
D^{(1)}(E/k_1,\Th_{0}/\Th_1)\ 
D^{(1)}(E/k_2,\Th_{0}/\Th_2)\,\nonumber\\
-\,\dt(\ln(E/k_1))\,\dt(\ln(E/k_2))
+\,\gamma_0^2\,\int_{k_1}^{E}\,\frac{dk}{k}
\,\int_{\Th_{12}}^{\Th_{0}}\,\frac{d\Th}{\Th}
\,D^{(2)}(k/k_1,k/k_2,\Th/\Th_1,\Th/\Th_2;\Th_{12}),\label{de}
\eqnx
}
where we have assumed (cf.\ (\ref{d2e}) ) that the intermediate emissions 
represented by the homogeneous term of (\ref{d2e}) 
{\it do not generate additional particle flow into either $\Th_1$ or $\Th_2$ 
window} ($\Th_{12}<\Th<\Th_0$). The
densities $D^{(1)}(E/k_1,\Th_{0}/\Th_1)$ and $D^{(1)}(E/k_2,\Th_{0}/\Th_2)$ 
(cf.\ Fig.\ 3a) denote here single particle inclusive densities for particles 
$k_1$ and $k_2$ respectively. All notations are as in Fig.\ 2.   
The density $D^{(2)}(E/k_1,E/k_2,\Th_{0}/\Th_{1}, \Th_{0}/\Th_{2};\Th_{12})$  
(cf.\ Fig.\ 3b) precisely represents (still in the  one-dimensional 
approximation) 
the inclusive two particle density to obtain two particles of energies 
$k_1$, $k_2$ from the subjet $(E, \Th_0)$ separated by the relative angle 
$\Th_{12}$. The ring apertures are $\Th_1$, $\Th_2$ respectively.
 
The relation between 
$D^{(2)}(E/k_1,E/k_2,\Th_{0}/\Th_{1},\Th_{0}/\Th_{2};\Th_{12})$ and the two 
particle density 
(\ref{fcqcd}) is the following~:
\eqn
D^{(2)}(P;\,E,\Th_0;\,k_1,k_2,\Th_{12},\Th_1,\Th_{2})=
D^{(1,ex)}_P(E,\Th_{0})
\cdot D^{(2)}(E/k_1,E/k_2,\Th_{0}/\Th_{1},\Th_{0}/\Th_{2};\Th_{12}),
\label{sumdiagr}
\eqnx
where $D^{(1,ex)}_P(E,\Th_{0})$ is the exclusive single particle density
to produce a subjet of energy $E$ placed at the opening angle $\Th_0$
with respect to the main jet $P$.
 
To sum up, convolution (\ref{sumdiagr}) represents the emission of 
particles from one subjet $(E,\Th_{0})$ originating from the main jet 
$(P,\Th_P)$ into two rings of apertures  $\Th_{1}, \Th_{2}$ with separation 
angle $\Th_{12}$, placed around the subjet axis (see Fig.\ 4).

Introducing new variables~:
\eqn
x_1=\frac{k_1}{E},\,
w_{12}=\frac{k_1}{k_2},\,
%\nonumber\\
y_2=\ln \frac{\Th_{0}}{\Th_{12}},\,
%&\nonumber\\
t_{1}=\ln\frac{\Th_{12}}{\Th_1},\,
%&& 
t_{2}=\ln\frac{\Th_{12}}{\Th_2},
\label{var}
\eqnx
we rewrite (\ref{d2e}) as~:
\eqn
D^{(2)}(1/x_1,w_{12},t_{1},t_2,y_2)&=&
D^{(1)}(1/x_1,y_2+t_{1})\,D^{(1)}(w_{12}/x_1,y_2+t_2)\nonumber\\
&-&\dt(\ln 1/x_1)\,\dt(\ln w_{12}/x_1)\,\label{d2b}\\
&+&\,\gamma_0^2\,\int_{x_1}^{1}\,\frac{dx_1^{\prime}}{x_1^{\prime}}\,
\int_{0}^{y_2}\,dy_2^{\prime}
\,D^{(2)}(1/x_1^{\prime},w_{12},t_{1},t_2,y_2^{\prime})
\nonumber.
\eqnx

In order to solve Eq.\ (\ref{d2b}), we transform it into moment space
by means of Mellin transform~:
\eq
D^{(2)}(\om,w_{12},t_{1},t_2,y_2)=\int_{0}^{1}\,dx_1\,x_1^{\om-1}\,
D^{(2)}(1/x_1,w_{12},t_{1},t_2,y_2).
\label{mellin}
\eqx
Then we differentiate both sides of (\ref{d2b}) with respect to $y_2$. 
Finally, we obtain~:
\eq
\frac{d}{dy_2} D^{(2)}(\om,w_{12},t_{1},t_{2},y_2)=
r(\om,w_{12},t_{1},t_{2},y_2)
+\frac{\gamma_0^2}{\om}\,D^{(2)}(\om,w_{12},t_{1},t_{2},y_2),
\label{d2rozn}
\eqx
where the function $r(\om,w_{12},t_{12},y_2)$ reads~:
\eq
r(\om,w_{12},t_{1},t_{2},y_2)=\int_{0}^{1}\,dx_1\,x_1^{\om-1}\,
D^{(1)}(1/x_1,y_2+t_{1})\,D^{(1)}(w_{12}/x_1,y_2+t_2).
\label{r}
\eqx
Equation (\ref{d2rozn}) is an ordinary inhomogeneous linear 
differential equation. Taking into account initial conditions defined by 
(\ref{d2b}), its explicit solution takes the form~:
\eqn
D^{(2)}(\om,w_{12},t_{1},t_{2},y_2)=r(\om,w_{12},t_{1},t_{2},y_2)\,\,+\frac
{\gamma_0^2}{\om}\,e^{\frac{\gamma_0^2}{\om}y_2}\,
R(\om,w_{12},t_{1},t_{2},y_2)\,\,\nonumber\\
-\dt(w_{12}-1)\,e^{\frac{\gamma_0^2}{\om}y_2} 
-\frac{\gamma_0^2}{\om}\,e^{\frac{\gamma_0^2}{\om}y_2}\,
R(\om,w_{12},t_{1},t_{2},0),
\label{d2expl0}
\eqnx
where function $r(\om,w_{12},t_{1},t_{2},y_2)$ was defined in 
(\ref{r}), and function $R(\om,w_{12},t_{1},t_{2},y_2)$ denotes
the following indefinite integral of $r(\om,w_{12},t_{12},y_2)$~:
\eq
R(\om,w_{12},t_{1},t_{2},y_2)=\int^{y_2} dy\, r(\om,w_{12},t_{1},t_{2},y)
e^{-\frac{\gamma_0^2}{\om}y_2}.
\label{rr}
\eqx
%
%%%%%%%%%%%%%%%%%%%%%%%%%%%%%%%%%%%%%%%%%%%%%%%%%%%%%%%%%%%%%%%%%%%%%%%%%%%%%%
%
%%%%%%%%%%%%%%%%%%%%%%%%%%%%%%%%%%%%%%%%%%%%%%%%%%%%%%%%%%%%%%%%%%%%%%%%%%%%%%
%

\section{The QCD factorial correlators: Derivation}

As a function of the  two-particle inclusive density
(\ref{d2expl0}),  the convolution 
of $D^{(2)}$ and multiplicity correlators in (\ref{fcqcd}) can be 
now expressed in explicit form. After a new change of
variables~:
\eqn
l_1=\log \frac{E}{k_1},& & s_{12}=\log \frac{k_1}{k_2}
\label{varls}
\eqnx
the equation (\ref{fcqcd}) rewritten for the {\it  normalized angular 
correlators} 
\eq
{\bar F}_{q_1\,q_2}(\Th_{0}/\Th_1,\Th_{0}/\Th_2)=
\frac{F_{q_1\,q_2}(\Th_{0};\Th_{12},\Th_{1},\Th_2)}
{D^{(1,ex)}_P(E,\Th_{0})\,N^{q_1}(E\Th_{0})\,N^{q_2}(E\Th_{0})}
\label{fnorm}
\eqx
takes the form~:
\eqn
{\bar F}_{q_1\,q_2}(\Th_{0}/\Th_1,\Th_{0}/\Th_2)\sim 
\int_{0}^{\infty} dl_1 \,\int_{0}^{\infty} ds_{12} 
\,\int_{-i\infty}^{+i\infty} 
\frac{d\om}{2\pi i}\,e^{\om l_1}
\,D^{(2)}(\om,w_{12},t_{1},t_{2},y_2)\nonumber\\
\nonumber\\
\times e^{-q_1 \ga (y_2+t_1+l_1)}\,e^{-q_2 \ga (y_2+t_2+l_1+s_{12})},
\label{nfcqcd}
\eqnx
where we substituted  $D^{(2)}(1/x_1,w_{12},t_{1},t_{2},y_2)$
by its inverse Mellin representation~:
\eq
D^{(2)}(1/x_1,w_{12},t_{1},t_{2},y_2)=\int_{-i\infty}^{+i\infty} 
\frac{d\om}{2\pi 
i}\,(1/x_1)^{\om}\,D^{(2)}(\om,w_{12},t_{1},t_{2},y_2).
\label{imellin}
\eqx

Now let us calculate term by term the contributions to  
convolution integral (\ref{nfcqcd}) coming from the various 
components of the modified inclusive two particle
distribution (\ref{d2expl0}) denoted (I), (II), (III) 
and (IV) as follows: 
\eqn
D^{(2)}(\om,w_{12},t_{1},t_{2},y_2)&=&r(\om,w_{12},t_{1},t_2,y_2)\,\,
\rule{1.7cm}{0cm}{\bf (I)}\nonumber\\
\nonumber\\
&+&\frac{\gamma_0^2}{\om}\,e^{\frac{\gamma_0^2}{\om}y_2}\,
R(\om,w_{12},t_{1},t_2,y_2)\,\,{\bf (II)}\nonumber\\
\nonumber\\
&-&\dt(w_{12}-1)\,e^{\frac{\gamma_0^2}{\om}y_2}\,\,\rule{1.9cm}{0cm}
 {\bf (III)}\nonumber\\
\nonumber\\
&-&\frac{\gamma_0^2}{\om}\,e^{\frac{\gamma_0^2}{\om}y_2}\,
R(\om,w_{12},t_{1},t_2,0)\,\,{\bf (IV)},
\label{d2expl}
\eqnx

These terms have the following Mellin representation~:
\eqn
(I)\rule{1.8cm}{0cm} r(\om,w_{12},t_{1},t_2,y_2)=&&
\int_{-i\infty}^{+i\infty} \frac{d\om_1}{2\pi i}\,
\int_{-i\infty}^{+i\infty} \frac{d\om_2}{2\pi i}\label{d21}\\
&&\frac{1}{\om-\om_1-\om_2}\nonumber\\
&&\exp\left(\,\ga^2 y_2(\frac{1}{\om_1}+\frac{1}{\om_2})\,+\om_2
s_{12}+\frac{\ga^2}{\om_1}t_1\,+\frac{\ga^2}{\om_2}t_2\right),\nonumber\\
&&\nonumber\\
&&\nonumber\\
(II)\,\,\frac{\gamma_0^2}{\om}\,e^{\frac{\gamma_0^2}{\om}y_2}\,
R(\om,w_{12},t_{1},t_2,y_2)=&&
\int_{-i\infty}^{+i\infty} \frac{d\om_1}{2\pi i}\,
\int_{-i\infty}^{+i\infty} \frac{d\om_2}{2\pi i}\label{d22}\\
&&\frac{1}{\om-\om_1-\om_2}\,\,
\frac{1}{\om-\frac{\om_1 \om_2}{\om_1+\om_2}}\,\,\frac{\om_1 
\om_2}{\om_1+\om_2}
\nonumber\\
&&\times \exp\left(\,\ga^2 y_2(\frac{1}{\om_1}+\frac{1}{\om_2})\,+\om_2 
s_{12}\,+
\frac{\ga^2}{\om_1}t_1\,+\frac{\ga^2}{\om_2}t_2\right),\nonumber\\
&&\nonumber\\
&&\nonumber\\
(III)\rule{1.9cm}{0cm}
\dt(w_{12}-1)\,e^{\frac{\gamma_0^2}{\om}y_2}=&&\dt(s_{12})
\,e^{\frac{\gamma_0^2}{\om}y_2},
\label{d23}\\   
&&\nonumber\\
&&\nonumber\\
(IV)\,\,\frac{\gamma_0^2}{\om}\,e^{\frac{\gamma_0^2}{\om}y_2}\,
R(\om,w_{12},t_{1},t_2,0)=&&
\int_{-i\infty}^{+i\infty} \frac{d\om_1}{2\pi i}\,
\int_{-i\infty}^{+i\infty} \frac{d\om_2}{2\pi i}\label{d24}\\
&&\,\frac{1}{\om-\om_1-\om_2}\,\,
\frac{1}{\om-\frac{\om_1 \om_2}{\om_1+\om_2}}\,\,\frac{\om_1 
\om_2}{\om_1+\om_2}
\nonumber\\
&&\times \exp\left(\,\frac{\ga^2}{\om} y_2 \,+\om_2 s_{12}\,+
\frac{\ga^2}{\om_1}t_1\,+\frac{\ga^2}{\om_2}t_2\right).
\nonumber
\eqnx

The contributions of (\ref{d21}), (\ref{d22}), (\ref{d23}), 
(\ref{d24}) to (\ref{nfcqcd}) may be evaluated using the multidimensional 
saddle point approximation. For first term (\ref{d21}) its  
convolution (\ref{nfcqcd}) takes the form~:
\eqn
{\bar F}_{q_1\,q_2}^{I}(\Th_{0}/\Th_1,\Th_{0}/\Th_2)&\sim&
\int_{0}^{\infty} dl_1 \,\int_{0}^{\infty} ds_{12} 
\,\int_{-i\infty}^{+i\infty} 
\frac{d\om}{2\pi i}\,e^{\om l_1}\,r(\om,w_{12},t_{1},t_2,y_2)\nonumber\\
&&\times
e^{-q_1 \ga (y_2+t_1+l_1)}\,e^{-q_2 \ga (y_2+t_2+l_1+s_{12})}=\nonumber\\
\nonumber\\
&=&\int_{0}^{\infty} dl_1 \,\int_{0}^{\infty} ds_{12} \,
\int_{-i\infty}^{+i\infty} \frac{d\om}{2\pi i}
\int_{-i\infty}^{+i\infty} \frac{d\om_1}{2\pi i}\,
\int_{-i\infty}^{+i\infty} \frac{d\om_2}{2\pi i}\nonumber\\
&&\frac{1}{\om-\om_1-\om_2} \nonumber\\
&&\times\exp\left(-q_1 \ga (y_2+t_1+l_1)-q_2 \ga
(y_2+t_2+l_1+s_{12})+\om\,l_1\right)\nonumber\\
&&\times\exp\left(\,\ga^2 y_2(\frac{1}{\om_1}+\frac{1}{\om_2})\,+\om_2
s_{12}+\frac{\ga^2}{\om_1}t_1\,+\frac{\ga^2}{\om_2}t_2\right).\nonumber\\
\label{saddle21}
\eqnx
After performing the integral over the $\om$-pole in (\ref{f21}) 
one obtains~:

{\footnotesize
\eqn
{\bar F}_{q_1\,q_2}^{I}(\Th_{0}/\Th_1,\Th_{0}/\Th_2)\sim 
\int_{0}^{\infty} dl_1 \,\int_{0}^{\infty} ds_{12} \,
\int_{-i\infty}^{+i\infty} \frac{d\om_1}{2\pi i}\,
\int_{-i\infty}^{+i\infty} \frac{d\om_2}{2\pi i}
\exp\left(\,S(\om_1,\om_2,l_1,s_{12},t_1,t_2,y_2)\right)
\label{f21}
\eqnx
}

\noindent
which in the saddle point approximation may be estimated as~:

\eqn
\int_{0}^{\infty} dl_1 \,\int_{0}^{\infty} ds_{12} \,
\int_{-i\infty}^{+i\infty} \frac{d\om_1}{2\pi i}\,
\int_{-i\infty}^{+i\infty} \frac{d\om_2}{2\pi i}
\exp\left(\,S(\om_1,\om_2,l_1,s_{12},t_1,t_2,y_2)\right)\sim\nonumber\\
\nonumber\\
\exp\left(\,S(\om_1,\om_2,l_1,s_{12},t_1,t_2,y_2)
\mid_{\frac{\pa S}{\pa l_1}=0,\, \frac{\pa S}{\pa s_{12}}=0,\,
\frac{\pa S}{\pa \om_1}=0,\,\frac{\pa S}{\pa \om_2}=0}\right)
\paus\nonumber\\
\times \,\, det \, \left(
\begin{array}{cccc}  
\frac{\pa^2 S}{\pa l_1^2}           & \frac{\pa^2 S}{\pa l_1 \pa s_{12}}& 
\frac{\pa^2 S}{\pa l_1 \pa \om_1}   & \frac{\pa^2 S}{\pa l_1 \pa \om_2} \\
\frac{\pa^2 S}{\pa s_{12} \pa l_1}  & \frac{\pa^2 S}{\pa s_{12}^2}       &
\frac{\pa^2 S}{\pa s_{12} \pa \om_1}& \frac{\pa^2 S}{\pa s_{12} \pa \om_2}\\ 
\frac{\pa^2 S}{\pa \om_1 \pa l_1}   & \frac{\pa^2 S}{\pa \om_1 \pa s_{12}}&
\frac{\pa^2 S}{\pa \om_1^2}         & \frac{\pa^2 S}{\pa \om_1 \pa \om_2}  
\\
\frac{\pa^2 S}{\pa \om_2 \pa l_1}   & \frac{\pa^2 S}{\pa \om_2 \pa s_{12}}&
\frac{\pa^2 S}{\pa \om_2 \pa \om_1} & \frac{\pa^2 S}{\pa \om_2^2}\\
\end{array} 
\right)^{-1/2}_{\frac{\pa S}{\pa l_1}=0,\, \frac{\pa S}{\pa s_{12}}=0,\,
\frac{\pa S}{\pa \om_1}=0,\,\frac{\pa S}{\pa \om_2}=0}.
\label{gsaddle}
\eqnx
For (\ref{f21}), the function 
$S^{I}(\om_1,\om_2,l_1,s_{12},t_{1},t_{2},y_2)$ reads~:
\eqn 
S^{I}(\om_1,\om_2,l_1,s_{12},t_1,t_2,y_2)&=&
-q_1 \ga (y_2+t_1+l_1)-q_2 \ga 
(y_2+t_2+l_1+s_{12})\label{s21}\\
&+&(\om_1+\om_2)l_1
+\ga^2 y_2(\frac{1}{\om_1}+\frac{1}{\om_2})\,+\om_2 s_{12}
+\frac{\ga^2}{\om_1}t_1\,+\frac{\ga^2}{\om_2}t_2.
\nonumber
\eqnx
Hence after evaluating (\ref{f21}) one obtains~:
\eqn
{\bar F}_{q_1\,q_2}^{I}(\Th_{0}/\Th_1,\Th_{0}/\Th_2)\sim \rule{10.5cm}{0cm}
\nonumber\\
-\frac{1}{4\pi^2}\,\exp \left\{ y_2\,\left(\am - \bm +\an -\bn \right)\right.\,
\left.+t_1\,\left(\am - \bm \right)\,+t_2\,\left(\an - \bn \right)\right\}.
\label{sf21}
\eqnx
For the second term (\ref{d22}), convolution (\ref{nfcqcd}) 
takes the form~:
\eqn
{\bar F}_{q_1\,q_2}^{II}(\Th_{0}/\Th_1,\Th_{0}/\Th_2)&\sim&
\int_{0}^{\infty} dl_1 \,\int_{0}^{\infty} ds_{12} 
\,\int_{-i\infty}^{+i\infty} 
\frac{d\om}{2\pi i}\,e^{\om l_1}\,
\frac{\gamma_0^2}{\om}\,e^{\frac{\gamma_0^2}{\om}y_2}\,
R(\om,w_{12},t_1,t_2,y_2) \nonumber\\
&&\times e^{-q_1 \ga (y_2+t_1+l_1)}\,e^{-q_2 \ga 
(y_2+t_2+l_1+s_{12})}=\nonumber\\
\nonumber\\
&=&\int_{0}^{\infty} dl_1 \,\int_{0}^{\infty} ds_{12} \,
\int_{-i\infty}^{+i\infty} \frac{d\om}{2\pi i}
\int_{-i\infty}^{+i\infty} \frac{d\om_1}{2\pi i}\,
\int_{-i\infty}^{+i\infty} \frac{d\om_2}{2\pi i}\nonumber\\
&&\frac{1}{\om-\om_1-\om_2}\,\,
\frac{1}{\om-\frac{\om_1 \om_2}{\om_1+\om_2}}\,\,\frac{\om_1 
\om_2}{\om_1+\om_2}
\nonumber\\
&&\times \exp\left(-q_1 \ga (y_2+t_1+l_1)-q_2 \ga 
(y_2+t_2+l_1+s_{12})+\om\,l_1\right)\nonumber\\
&&\times \exp\left(\,\ga^2 y_2(\frac{1}{\om_1}+\frac{1}{\om_2})\,+\om_2 
s_{12}\,+
\frac{\ga^2}{\om_1}t_1\,+\frac{\ga^2}{\om_2}t_2\right).\nonumber\\
\label{f22}
\eqnx
Since there are two $\om$-poles in (\ref{f22}) the integration over 
$\om$ will give rise to two separate saddle point exponents~:
{\footnotesize
\eqn
\int_{-i\infty}^{+i\infty}\frac{d\om}{2\pi i}\,e^{\om l_1}
\frac{1}{\om-\om_1-\om_2}\,\,
\frac{1}{\om-\frac{\om_1 \om_2}{\om_1+\om_2}}\,\,\frac{\om_1 
\om_2}{\om_1+\om_2}
=\left(e^{(\om_1+\om_2)l_1}-e^{\frac{\om_1 
\om_2}{\om_1+\om_2}l_1}\right)
\frac{\om_1 \om_2}{\om_1^2+\om_1 \om_2 +\om_2^2}
\label{rozw22}
\eqnx
}
which have to be evaluated separately. One obtains~:
\eqn
{\bar F}_{q_1\,q_2}^{II}(\Th_{0}/\Th_1,\Th_{0}/\Th_2)\sim\int_{0}^{\infty} dl_1 
\,\int_{0}^{\infty} ds_{12} \,
\int_{-i\infty}^{+i\infty} \frac{d\om_1}{2\pi i}\,
\int_{-i\infty}^{+i\infty} \frac{d\om_2}{2\pi i}
\label{saddle22}\rule{3cm}{0cm}\\
\left\{ \exp(\,S^{IIa}(\om_1,\om_2,l_1,s_{12},t_1,t_2,y_2))\right.
-\left.\exp(\,S^{IIb}(\om_1,\om_2,l_1,s_{12},t_1,t_2,y_2))\right\},
\nonumber\rule{1cm}{0cm}
\eqnx
where $S^{IIa}$, $S^{IIb}$ read~:
\eqn 
S^{IIa}(\om_1,\om_2,l_1,s_{12},t_1,t_2,y_2)&=&
-q_1 \ga (y_2+t_1+l_1)-q_2 \ga 
(y_2+t_2+l_1+s_{12})\nonumber\\
&+&(\om_1+\om_2)l_1+\ga^2y_2(\frac{1}{\om_1}+\frac{1}{\om_2})\,+\om_2 s_{12}
\nonumber\\
&+&\frac{\ga^2}{\om_1}t_1\,+\frac{\ga^2}{\om_2}t_2 
+\ln \frac{\om_1 \om_2}{\om_1^2+\om_1 \om_2 +\om_2^2}, 
\label{s22a}
\eqnx
\eqn 
S^{IIb}(\om_1,\om_2,l_1,s_{12},t_1,t_2,y_2)&=&
-q_1 \ga (y_2+t_1+l_1)-q_2 \ga (y_2+t_2+l_1+s_{12})
\nonumber\\
&+&\frac{\om_1\om_2}{\om_1+\om_2} l_1+\ga^2 y_2(\frac{1}{\om_1}
+\frac{1}{\om_2})\,+\om_2 s_{12}\nonumber\\
&+&\frac{\ga^2}{\om_1}t_1\,+\frac{\ga^2}{\om_2}t_2
+\ln \frac{\om_1 \om_2}{\om_1^2+\om_1 \om_2 +\om_2^2}. 
\label{s22b}
\eqnx
Hence, after evaluating (\ref{f22}) one obtains~: 

{\footnotesize
\eqn
{\bar F}_{q_1\,q_2}^{II}(\Th_{0}/\Th_1,\Th_{0}/\Th_2)\sim
\label{sf22}\rule{11.5cm}{0cm}\\
-\frac{1}{4\pi^2} \frac{q_1 q_2}{q_1^2+q_1 q_2 +q_2^2}
\left\{ \exp \left[ y_2\left(\am - \bm +\an -\bn \right)
+t_1\left(\am - \bm \right)
+t_2\left(\an - \bn \right)\right]\right.\nonumber\\
\left.-\frac{(q_1+q_2)q_2}{q_1^2}\,\exp[ y_2\left(\amn - \bmn \right)
+t_1\left(-\bm - \frac {\bm}{q_2(q_1+q_2)}\left. \right)
+t_2\left(\an - \bn \right)\right] \right\}.
\nonumber
\eqnx
}

\noindent
Similarly, for the third  (\ref{d23}) and fourth term (\ref{d24}),  
their convolutions (\ref{nfcqcd}) result in~:
\eqn
{\bar F}_{q_1\,q_2}^{III}(\Th_{0}/\Th_1,\Th_{0}/\Th_2)&\sim&
\int_{0}^{\infty} dl_1 \,\int_{0}^{\infty} ds_{12} 
\,\int_{-i\infty}^{+i\infty} 
\frac{d\om}{2\pi i}\,e^{\om l_1}\,
\dt(s_{12})\,e^{\frac{\gamma_0^2}{\om}y_2}\nonumber\\
&&\times e^{-q_1 \ga (y_2+t_1+l_1)}\,e^{-q_2 \ga 
(y_2+t_2+l_1+s_{12})}=\nonumber\\
\nonumber\\
&=&\int_{0}^{\infty} dl_1 \,\int_{-i\infty}^{+i\infty} \frac{d\om}{2\pi i}
\,\int_0^{\infty}ds_{12}\,\int_{-i\infty}^{+i\infty}\frac{dp}{2\pi i}
\nonumber\\
&&\times\exp\left(-q_1 \ga (y_2+t_1+l_1)-q_2 \ga 
(y_2+t_2+l_1+s_{12})+\om\,l_1\right)\nonumber\\
&&\times\exp\left(\frac{\gamma_0^2}{\om}y_2+p s_{12}\right),
\label{f23}
\eqnx
\eqn
{\bar F}_{q_1\,q_2}^{IV}(\Th_{0}/\Th_1,\Th_{0}/\Th_2)&\sim&
\int_{0}^{\infty} dl_1 \,\int_{0}^{\infty} ds_{12} 
\,\int_{-i\infty}^{+i\infty} 
\frac{d\om}{2\pi i}\,e^{\om l_1}\,
\frac{\gamma_0^2}{\om}\,e^{\frac{\gamma_0^2}{\om}y_2}\,
R(\om,w_{12},t_{1},t_2,0)\nonumber\\
&&\times e^{-q_1 \ga (y_2+t_1+l_1)}\,e^{-q_2 \ga 
(y_2+t_2+l_1+s_{12})}=\nonumber\\
\nonumber\\
&=&\int_{0}^{\infty} dl_1 \,\int_{0}^{\infty} ds_{12} \,
\int_{-i\infty}^{+i\infty} \frac{d\om}{2\pi i}
\int_{-i\infty}^{+i\infty} \frac{d\om_1}{2\pi i}\,
\int_{-i\infty}^{+i\infty} \frac{d\om_2}{2\pi i}\nonumber\\
&&\frac{1}{\om-\om_1-\om_2}\,\,
\frac{1}{\om-\frac{\om_1 \om_2}{\om_1+\om_2}}\,\,\frac{\om_1 
\om_2}{\om_1+\om_2}
\nonumber\\
&&\times\exp\left(-q_1 \ga (y_2+t_1+l_1)-q_2 \ga 
(y_2+t_2+l_1+s_{12})+\om\,l_1\right)\nonumber\\
&&\times\exp\left(\,y_2\frac{\ga^2}{\om}\,+\om_2 s_{12}\,+
\frac{\ga^2}{\om_1}t_1\,+\frac{\ga^2}{\om_2}t_2\right).\nonumber\\
\label{f24}
\eqnx
The respective saddle functions obtained similarly as in 
(\ref{rozw22}) and (\ref{saddle22}) are the following~:
\eqn 
S^{III}(\om,p,l_1,s_{12},t_1,t_2,y_2)&=&
-q_1 \ga (y_2+t_1+l_1)-q_2 \ga (y_2+t_2+l_1+s_{12})\nonumber\\
&+&\om l_1\,+ \frac{\ga^2}{\om}y_2+p s_{12},
\label{s23}\\
\nonumber\\
S^{IVa}(\om_1,\om_2,l_1,s_{12},t_1,t_2,y_2)&=&
-q_1 \ga (y_2+t_1+l_1)-q_2 \ga 
(y_2+t_2+l_1+s_{12})\nonumber\\
&+&(\om_1+\om_2)l_1+\ga^2 y_2 \frac{1}{\om_1+\om_2}\,+\om_2 s_{12}\nonumber\\
&+&\frac{\ga^2}{\om_1}t_1\,+\frac{\ga^2}{\om_2}t_2 
+\ln \frac{\om_1 \om_2}{\om_1^2+\om_1 \om_2 +\om_2^2}, 
\label{s24a}\\
\nonumber\\
S^{IVb}(\om_1,\om_2,l_1,s_{12},t_1,t_2,y_2)&=&
-q_1 \ga (y_2+t_1+l_1)-q_2 \ga (y_2+t_2+l_1+s_{12})\nonumber\\
&+&\frac{\om_1 \om_2}{\om_1+\om_2} l_1+\ga^2 y_2(\frac{1}{\om_1}
+\frac{1}{\om_2})\,+\om_2 s_{12}\nonumber\\
&+&\frac{\ga^2}{\om_1}t_1\,+\frac{\ga^2}{\om_2}t_2
+\ln \frac{\om_1 \om_2}{\om_1^2+\om_1 \om_2 +\om_2^2}. 
\label{s24b}
\eqnx
Hence after evaluating (\ref{f23}),(\ref{f24})  one obtains~: 
{\footnotesize
\eqn
{\bar F}_{q_1\,q_2}^{III}(\Th_{0}/\Th_1,\Th_{0}/\Th_2)&\sim& 
-\frac{1}{4\pi^2}\,\exp \left\{ \right. y_2\,\left(\amn - \bmn \right)\,
-t_1\,\left( \bm \right)\,
-t_2\,\left( \bn \right)\left. \right\},
\label{sf23}
\eqnx
}
{\footnotesize
\eqn
{\bar F}_{q_1\,q_2}^{IV}(\Th_{0}/\Th_1,\Th_{0}/\Th_2)\sim
\label{sf24}\rule{11.5cm}{0cm}\\
-\frac{1}{4\pi^2}\frac{q_1 q_2}{q_1^2+q_1 q_2 +q_2^2}
\left\{\exp \left[y_2\left(\amn - \bmn \right)\right.
+t_1\left(\am - \bm \right)
+t_2\left(\an - \bn \right)\right]
\nonumber\\
-\frac{(q_1+q_2)q_2}{q_1^2} 
\exp \left[ y_2\left(\amn - \bmn \right)\right.
+t_1\left(-\bm - \frac {\bm}{q_2(q_1+q_2)} \right)
\left.\left.+t_2\left(\an - \bn \right)\right] \right\}.
\nonumber
\eqnx
}

We now have to sum up contributions (\ref{sf21}), (\ref{sf22}), 
(\ref{sf23}), (\ref{sf24}) according to  (\ref{nfcqcd}), (\ref{d2expl}). 
Hence, finally, {\it normalized angular correlators} (\ref{fnorm}) read~:
\eqn
{\bar F}_{q_1\,q_2}(\Th_{0}/\Th_1,\Th_{0}/\Th_2)&\sim& 
\frac{1}{4\pi^2}\,A\,e^{y_2\,\left(\amn - \bmn \right)\,
+t_1\,\left(\am - \bm \right)\,+t_2\,\left(\an - \bn \right)}\label{sf}\\
&+&\frac{1}{4\pi^2}\,e^{y_2\,\left(\amn - \bmn \right)\,
-t_1\,\left( \bm \right)\,-t_2\,\left( \bn \right)}\nonumber\\
&-&\frac{1}{4\pi^2}\,(A+1)\,e^{ y_2\,\left(\am - \bm +\an -\bn \right)\,
+t_1\,\left(\am - \bm \right)\,+t_2\,\left(\an - \bn \right)},\nonumber
\eqnx
where $A=\frac{q_1q_2}{q_1^2+q_1q_2+q_2^2}$.

After dividing (\ref{sf}), in analogy to (\ref{fq12scal}), by the product 
$F_{q_1}(\Th_{0}/\Th_1)\,F_{q_2}(\Th_{0}/\Th_2)$ 
one obtains~:
\eqn
\frac{{\bar F}_{q_1\,q_2}(\Th_{0}/\Th_1,\Th_{0}/\Th_2)}
{F_{q_1}(\Th_{0}/\Th_1)\,F_{q_2}(\Th_{0}/\Th_2)}&\sim&
\frac{1}{4\pi^2}\,A
\left(\frac{\Th_{0}}{\Th_{12}}\right)^{\phi_{q_1+q_2}-\phi_{q_1}-\phi_{q_2}}
-\frac{1}{4\pi^2}(A+1)\nonumber\\
&+&\frac{1}{4\pi^2}
\left(\frac{\Th_{0}}{\Th_{12}}\right)^{\phi_{q_1+q_2}-\phi_{q_1}-\phi_{q_2}}
\left(\frac{\Th_1}{\Th_{12}}\right)^{\gamma_0/q_1}\,
\left(\frac{\Th_2}{\Th_{12}}\right)^{\gamma_0/q_2},\label{sff}
\eqnx
where $\phi_{q_1+q_2}=\gamma_0/(q_1+q_2)-\gamma_0(q_1+q_2)$ and 
$\phi_{q_1}=\gamma_0/q_1-\gamma_0\,q_1$, 
$\phi_{q_2}=\gamma_0/q_2-\gamma_0\,q_2$.  Note that the quantity 
$\phi_{q_1+q_2}-\phi_{q_1}-\phi_{q_2}$ has been denoted $\phi_{q_1,\,q_2}$ 
in formula (\ref{expo}).

%%%%%%%%%%%%%%%%%%%%%%%%%%%%%%%%%%%%%%%%%%%%%%%%%%%%
\section{Angular scaling of factorial correlators}
%%%%%%%%%%%%%%%%%%%%%%%%%%%%%%%%%%%%%%%%%%%%%%%%%%%%

In order to discuss the physical properties of QCD  factorial correlators, let 
us rewrite formula (\ref{sf}) in a different form
\eqn
{\bar F}_{q_1\,q_2}(\Th_{0}/\Th_1,\Th_{0}/\Th_2)\sim\rule{10.5cm}{0cm}
\nonumber\\
\frac{1}{4\pi^2}\,A\ 
\left\{\left(\frac{\Th_{0}}{\Th_{12}}\right)^{\phi_{q_1+q_2}}
\left(\frac{\Th_{12}}{\Th_1}\right)^{\phi_{q_1}}\,
\left(\frac{\Th_{12}}{\Th_2}\right)^{\phi_{q_2}} - 
\left(\frac{\Th_{0}}{\Th_{1}}\right)^{\phi_{q_1}}
\left(\frac{\Th_{0}}{\Th_{2}}\right)^{\phi_{q_2}}
\right\}\label{physics}
\nonumber\\
+
\frac{1}{4\pi^2}\ 
\left\{\left(\frac{\Th_{0}}{\Th_{12}}\right)^{\phi_{q_1+q_2}}
\left(\frac{\Th_{12}}{\Th_1}\right)^{\phi_{q_1}-\gamma_0/q_1}\,
\left(\frac{\Th_{12}}{\Th_2}\right)^{\phi_{q_2}-\gamma_0/q_2} - 
\left(\frac{\Th_{0}}{\Th_{1}}\right)^{\phi_{q_1}}
\left(\frac{\Th_{0}}{\Th_{2}}\right)^{\phi_{q_2}}
\right\}\ .
\eqnx
The physical interpretation of the two brackets contributing to formula 
(\ref{physics}) is quite simple. Considering the first one which is dominant at 
large  values of $\frac{\Th_{12}}{\Th_1}, \frac{\Th_{12}}{\Th_2}$ (\ref{ineq}), 
it corresponds to the contribution  coming from the full 
development of the parton cascade (minus the value when $\Th_{12}\equiv\Th_{0}$, 
i.e. substracting the effect of cascading {\it before} $\Th_{0}$). Indeed, due 
to the QCD constraints of angular ordering, the angular ordered path  
$\Th_{0}\rightarrow\Th_{12}$ is populated by fluctuations with order $q_1+q_2,$ 
while the remaining separated paths from $\Th_{12}\rightarrow\Th_{1}$ and 
$\Th_{12}\rightarrow\Th_{2}$ corresponds to the individual fluctuation patterns 
with order $q_1$ and $q_2.$ In QCD framework at DLA, this contribution is 
similar   to the  behaviour of the random cascading models. 

It is clear that this first term in equation (\ref{physics}) implies specific  
angular scaling properties of QCD jets (at DLA). Normalizing this term by 
the product $F_{q_1}(\Th_{0}/\Th_1)\,F_{q_2}(\Th_{0}/\Th_2)$  gives~:
\eqn
\frac{{\bar F}_{q_1\,q_2}(\Th_{0}/\Th_1,\Th_{0}/\Th_2)}
{F_{q_1}(\Th_{0}/\Th_1)\,F_{q_2}(\Th_{0}/\Th_2)}&\sim&
\frac{1}{4\pi^2}\,A
\left(\frac{\Th_{0}}{\Th_{12}}\right)^{\phi_{q_1+q_2}-\phi_{q_1}-\phi_{q_2}}
\label{sfff}\ .
\eqnx
The scaling properties of (\ref{sfff}) can be expressed by the following  three 
items

i) The normalized correlator (at DLA) depends only on the angular separation  
$\Th_{12},$ and thus is independent of the window sizes $\Th_1,\Th_2,$ 

ii) It obeys a scaling law as a function of the ratio 
$\frac{\Th_{0}}{\Th_{12}}$, 

iii) The scaling exponent  is related to the ones of the factorial moments by 
$\phi_{q_1,q_2}= \phi_{q_1+q_2}-\phi_{q_1}-\phi_{q_2}$.

Such a prediction is similar to the one of random cascading models, which has 
previously \cite{l16} been discussed for soft hadronic multiproduction. In that 
case, the property i) has been verified, while the dependence ii) showed some 
bending and iii) was largely violated since from observation in some range of 
resolution it appeared that $\phi_{q_1,q_2}\gg \phi_{q_1+q_2}-\phi_{q_1}
-\phi_{q_2}$. 
We know now that the multiplicity fluctuations in soft hadronic multiproduction
are influenced by Bose-Einstein enhancements. It would thus be interesting to 
measure by comparison the normalized correlators in jets, where the dynamics 
is more directly related to perturbative (and resummed) QCD properties.
The experimental analysis can be done as an extension of what was done for
angular factorial moments \cite{l8}, where window rings around the jet axis
have been considered as phase-space slices. 

Interestingly enough, a second contribution appears in formula (\ref{physics}) 
which also has a simple physical interpretation. On the contrary with the first 
term, the exponents $\phi_{q_1}-\gamma_0/q_1$ and $\phi_{q_1}-\gamma_0/q_1$ mean 
that the parton cascading structure during the second step of the process 
related to the separated paths from $\Th_{12}\rightarrow\Th_{1}$ and 
$\Th_{12}\rightarrow\Th_{2}$  is damped, since the corresponding fractal 
dimensions  $\gamma_0/q_1$ and $\gamma_0/q_1$ are cancelled from the 
intermittency exponents. This corresponds to the probability of having a  
contribution of parton jets directly into the windows of observation. This 
contribution is obviously subdominant at DLA, since the exponents are smaller. 
It would lead to a violation of the scaling properties i)-iii).

However, it remains to be found whether,  beyond the  DLA approximation, such a 
contribution could be in practice  larger than  the first one. In particular, 
the lack of DLA  exponentiation could be compensated by the strong contribution
to multiplicity fluctuations of subjets  directly hitting the observation 
windows. This study is beyond the scope of our paper devoted to the analysis 
of the DLA approximation but is deserved in the future. 

%%%%%%%%%%%%%%%%%%%%%%%%%%%%%%%%%%%%%%%%%%%%%%%%%%%%%%%%%
\section{Summary} 
%%%%%%%%%%%%%%%%%%%%%%%%%%%%%%%%%%%%%%%%%%%%%%%%%%%%%%%%%

We presented the analytical derivation of factorial correlators performed 
for the QCD parton cascade at the double logarithmic (DL) accuracy. 
For simplicity we considered only the fixed $\alpha_S$ case, expecting that 
it gives good qualitative estimation of scaling exponents as it was realized 
previously for factorial moments.
The scaling dependence of the correlators on the relative distance
between the two solid-angle cells recovered the similar result obtained
in the framework of random cascading $\alpha-$model \cite{l2,l6}, and seems to 
be a kind of universal relation.

However, it remains to be found whether the scaling holds also beyond the DLA 
approximation, where the contribution to multiplicity fluctuations coming 
from subjets  directly hitting the observation windows may be dominant.
This study is beyond the scope of our paper devoted to the analysis of the 
DLA approximation but is deserved in the future.

It would also be useful to compare these predictions with QCD Monte-Carlo 
simulations (based on parton showers). It is already known that there is a 
noticeable difference between QCD  predictions at DLA and QCD Monte-Carlo 
simulations for factorial moments, these predictions being in better agreement 
(but  not perfect) with data. Since the origin of this discrepancy is not well 
understood at present, the study of factorial correlators could be useful for 
identifying the problem.

%%%%%%%%%%%%%%%%%%%%%%%%%%%%%%%%%%%%%%%%%%%%%%%%%%%%%%%%%%%%%%%%%%%%%%%%%%
\section*{Acknowledgments}

Authors would like to thank B. Buschbeck for discussions,
G. Ingelman for reading the manuscript and useful comments. 
B.\ Z.\ would like to thank the members of 
CEA, Service de Physique Th\'eorique for their kind hospitality
during her stay in Saclay.
This research has been supported in part by the Polish Committee 
for Scientific Research grant Nos.\ 2 P03B 04718, 2 P03B 05119, 
European Community grant "Training and Mobility of Researchers", 
Network "Quantum Chromodynamics and the Deep Structure of Elementary 
Particles" FMRX-CT98-0194 and Subsydium 1/99 of the Foundation for Polish
Science. B.\ Z.\ is a fellow of the Stefan Batory Foundation.
%%%%%%%%%%%%%%%%%%%%%%%%%%%%%%%%%%%%%%%%%%%%%%%%%%%%%%%%%%%%%%%%%
%%%%%%%%%
%

\end{document}